\documentclass[prb,amsmath,amssymb,superscriptaddress,twocolumn]{revtex4}
\usepackage{float}
\usepackage{amsmath}
\usepackage{amssymb}
\usepackage{amsfonts}
\usepackage{euscript}
\usepackage{enumerate}
\usepackage{hhline}
\usepackage{pslatex}
\usepackage{tabularx}
\usepackage[usenames,dvipsnames]{xcolor}

\usepackage{graphicx}

\usepackage{dcolumn}
\usepackage{bm}
\usepackage[sort&compress]{natbib}

\usepackage{lipsum}

\makeatletter
\renewcommand{\@biblabel}[1]{#1. }
\renewcommand{\@dotsep}{500}
\renewcommand{\@pnumwidth}{0em}
\renewcommand{\l@figure}[2]{
\@dottedtocline{1}{1.5em}{2em}{Figure #1}{}\vspace{15pt}}

\newcommand{\ks}[1]{\textcolor{black}{#1}}
\newcommand{\kss}[1]{\textcolor{black}{#1}}
\newcommand{\xl}[1]{\textcolor{black}{#1}}

\newcommand{\wmk}[1]{\textcolor{black}{#1}}
\newcommand{\reply}[1]{\textcolor{black}{#1}}

\begin{document}
\title{Fractional optical angular momentum and multi-defect-mediated mode re-normalization and orientation control in photonic crystal microring resonators}

\author{Mingkang Wang} 
\affiliation{Microsystems and Nanotechnology Division, Physical Measurement Laboratory, National Institute of Standards and Technology, Gaithersburg, MD 20899, USA}
\affiliation{Department of
Chemistry and Biochemistry, University of Maryland, College Park, MD 20742, USA}
\author{Feng Zhou} 
\affiliation{Microsystems and Nanotechnology Division, Physical Measurement Laboratory, National Institute of Standards and Technology, Gaithersburg, MD 20899, USA}
\affiliation{Joint Quantum Institute, NIST/University of Maryland, College Park, MD 20742, USA}
\author{Xiyuan Lu}\email{xiyuan.lu@nist.gov}
\affiliation{Microsystems and Nanotechnology Division, Physical Measurement Laboratory, National Institute of Standards and Technology, Gaithersburg, MD 20899, USA}
\affiliation{Joint Quantum Institute, NIST/University of Maryland, College Park, MD 20742, USA}
\author{Andrew McClung}
\affiliation{Department of Electrical and Computer Engineering, University of
Massachusetts Amherst, Amherst, MA 01003, USA}
\author{Marcelo Davanco}
\affiliation{Microsystems and Nanotechnology Division, Physical Measurement Laboratory, National Institute of Standards and Technology, Gaithersburg, MD 20899, USA}
\author{Vladimir A. Aksyuk}
\affiliation{Microsystems and Nanotechnology Division, Physical Measurement Laboratory, National Institute of Standards and Technology, Gaithersburg, MD 20899, USA}
\author{Kartik Srinivasan} \email{kartik.srinivasan@nist.gov}
\affiliation{Microsystems and Nanotechnology Division, Physical Measurement Laboratory, National Institute of Standards and Technology, Gaithersburg, MD 20899, USA}
\affiliation{Joint Quantum Institute, NIST/University of Maryland, College Park, MD 20742, USA}

\date{\today}

\begin{abstract}
\noindent Whispering gallery modes (WGMs) in circularly symmetric optical microresonators exhibit integer quantized angular momentum numbers due to the boundary condition imposed by the geometry. Here, we show that incorporating a photonic crystal pattern in an integrated microring can result in WGMs with fractional optical angular momentum. By choosing the photonic crystal periodicity to open a photonic bandgap with a band-edge momentum lying between that of two WGMs of the unperturbed ring, we observe hybridized WGMs with half-integer quantized angular momentum numbers ($m \in \mathbb{Z}$ + 1/2). Moreover, we show that these modes with fractional angular momenta exhibit high optical quality factors with good cavity-waveguide coupling and an order of magnitude reduced group velocity. \wmk{\ks{Additionally,} by introducing multiple artificial defects, multiple modes can be localized to small volumes within the ring, \kss{while the relative orientation of the de-localized band-edge states can be well-controlled.}} Our work unveils \wmk{\ks{the renormalization of WGMs by} the photonic crystal, demonstrating novel fractional angular momentum states and nontrivial multi-mode \kss{orientation control} arising from continuous rotational symmetry breaking. The findings are expected to be useful for sensing/metrology, nonlinear optics, and cavity quantum electrodynamics.}
\end{abstract}  

\maketitle

Optical microresonators supporting whispering gallery modes (WGMs) have long been studied due to their ability to simultaneously support long cavity photon lifetime (high quality factor $Q$) and a strong degree of spatial mode localization, \ks{attributes that enhance light-matter interactions in areas such as cavity optomechanics~\cite{Aspelmeyer_RevModPhys_2014}, integrated frequency combs~\cite{shen2020integrated}, and  microlasers~\cite{rakovich2010photonic}}. WGMs have rotational symmetry in a circular round trip and, within a given wavelength band, have almost constant free spectral range (FSR). The circular geometry dictates the behavior of the electromagnetic field in the azimuthal ($\phi$) direction, as matching the field phase across one round trip (i.e, a $\phi$~=~2$\pi$ rotation) results in integer azimuthal mode numbers. Perturbations that disturb the geometry from the circular symmetry change this picture. For example, in deformed microcavities, geometric ray optics predicts more generalized WGMs with fractional angular momentum numbers of $m$/$n$~\cite{roll1998ray,Chang_OE_2017}, that is, the optical field exhibits $m\times 2\pi$ phase oscillations across $n$ round trips. Recently, such \kss{fractional angular momentum has} been demonstrated in a quadrupolar microdisk~\cite{Wang_LSA_2021}, in contrast to numerous free-space approaches that have been developed because of its potential applications in optical trapping, micromanipulation, and providing extra channels for multiplexing ~\cite{Beijersbergen_OptCommun_1994, Molina-Terriza_PRL_2002,Gotte_OE_2008,Gutierrez-Vega_JOA_2008, Nugrowati_OE_2012,Nugrowati_OC_2013, Ballantine_SciAdv_2016,Turpin_SR_2017}. However, the deformed microcavity geometry caused significant $Q$ deterioration, limiting the microcavity's ability to enhance light-matter interactions while exhibiting fractional angular momentum.

\ks{Conventional WGMs have also been modified through the introduction of artificial scatterers with designated size and position, in order to efficiently tune intermodal coupling and loss~\cite{peng2016chiral,wang2020electromagnetically}.The incorporation of a well-designed photonic crystal can realize coherent scattering between coupled modes and avoid coupling to radiation loss channels~\cite{Xu_APL_2019}. Recently, a `microgear' photonic crystal microring (MPhCR) has been developed~\cite{Lu_NatPhoton_2022}, in which clockwise (CW) and  counter-clockwise (CCW) traveling WGMs are hybridized into a set of standing WGMs through backscattering. This geometry enables strong intermodal coupling between WGMs with different angular momentum, resulting in large bandgaps (several FSRs) without decreasing the optical $Q$. As in other prior PhCR work~\cite{Lee_OL_2012, Zhang_PTL_2015, Gao_SciRep_2016, KML_OE_2017, Lo_OL_2018}, however, its photonic crystal structure only hosted conventional WGMs, with integer angular momentum, leaving WGMs with fractional angular momentum unexplored. Moreover, the incorporation of multiple defects and their influence on distributed modes within the ring was not considered.} \xl{Such understanding of the multiple defects/mode localization is crucial for utilizing the PhCR platform for cavity quantum electrodynamics.}

In this work, we show that the MPhCR structure can be modified to realize previously unexplored regimes of modal coupling in WGM resonators. First, we report fractional angular momentum WGMs in a MPhCR. By creating a band gap much larger than the FSR, we successfully hybridize two sets of WGMs with odd and even angular momentum, \kss{resulting} in WGMs of fractional angular momentum, while preserving the high $Q$, good coupling, and intuitive design common to conventional WGM resonators. Next, we explore a novel multi-mode localization process by introducing multiple artificial defects into the MPhCR. Breaking the continuous rotational symmetry of the microring geometry, the orientation and degeneracy of hybridized modes can be well controlled. Our work provides understanding of the modes' renormalization and \kss{orientation control} near the bandedge. It also highlights unique features of the PhCR, namely, the co-existence and interplay of the different types of states it supports.

\begin{figure}[t!]
\includegraphics[width=0.85\linewidth]{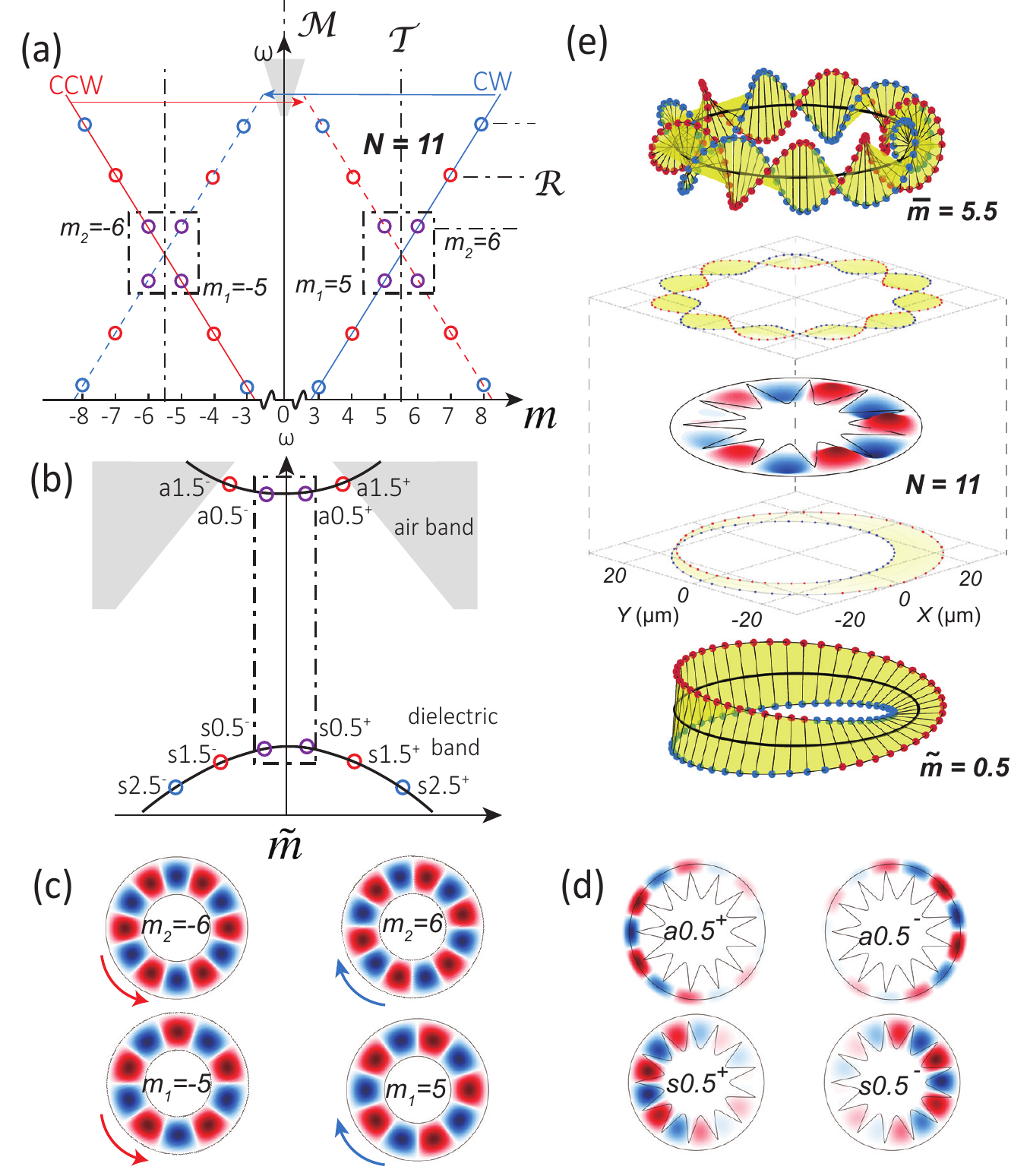}
\caption{\textbf{Fractional angular momentum whispering gallery modes enabled by mode superposition near a bandgap.}  \textbf{a,} The combination of rotational symmetry ($\mathcal{R}$) and period translation symmetry ($\mathcal{T}$) in a photonic crystal ring hybridizes the modes $|m_1+m_2| = N$, forming WGMs with fractional angular momentum when $N$ is odd. For $N=11$, the first order mode is formed by $|m_1|=5$ and $|m_2|=6$. Counterclockwise modes ($m<0$) are mirrored ($\mathcal{M}$) about the $y$-axis of $m = 0$. \textbf{b,} Band diagram of a photonic crystal ring with an odd number of unit cells. \wmk{Air/dielectric band modes denote modes overlapping largely with the air/dielectric part of the microring.} In a real experiment, the degenerate dielectric (air) modes split, showing higher and lower frequencies. We label them with $s(a)\tilde{m}^{+}$ and $s(a)\tilde{m}^{-}$, respectively. \textbf{c-d,} Qualitative illustrations of the four WGMs before (c) and after (d) hybridization. They correspond to the four purple circles framed by the dashed lines in (a) and (b), respectively.  ~\textbf{e,} An example of the modes with M{\"o}bius-type profile for both angular momentum phase (upper panel) and envelope phase (lower panel). The black circle aligns with the light propagation path.} 
\label{Fig1}
\end{figure}
For  a  microring  with  continuous rotational symmetry, the dominant traveling electric field of WGMs  can be written as $E(r,z,\phi) = E_{m}(r,z)~e^{im\phi}$, where $r$, $z$, $\phi$ are the radial, vertical, and azimuthal angular coordinate. $m$ represents the azimuthal angular momentum number of the mode, and matching the field phase across one round trip necessitates $m$ being \kss{a} positive or negative integer, corresponding to the CW and CCW WGMs. Figure~\ref{Fig1}(a) shows a schematic of the dispersion relation in a conventional microring. Over a certain range of frequencies, the WGMs have nearly equal FSR and symmetric CCW/CW modes due to the rotational symmetry ($\mathcal{R}$) and mirror symmetry ($\mathcal{M}$) of the microring. 


When $N$ azimuthally periodic modulations are introduced on the microring, it possesses translation symmetry ($\mathcal{T}$) (in the azimuthal direction) with a period of $2\pi/N$ and becomes a PhCR. The translation symmetry hybridizes four modes that exist near the bandgap, with angular momentum $|m_1+m_2|=N$ ~\cite{Lu_NatPhoton_2022}, generating dielectric and air bands, as shown in Fig.~\ref{Fig1}(b). To facilitate understanding, we consider that the four constituent traveling wave modes $\{|m_1|, -|m_1|, |m_2|, -|m_2|\}$ prior to mixing via the PhC are in-phase:

\vspace{-0.2 in}
\begin{equation}
\begin{split}
&E_{m_1}(r,z)~(e^{i|m_1|\phi} + e^{-i|m_1|\phi}) +  E_{m_2}(r,z)~(e^{i|m_2|\phi} + e^{-i|m_2|\phi})\\
& = ~~ E_{\bar{m}}(r,z) \cos(\bar{m}\phi)\cos(\tilde{m}\phi)
\end{split}\label{Eq1}
\end{equation}

\noindent where $E_{m_1}(r,z)$ and $E_{m_2}(r,z)$ are considered to be equal to $E_{\bar{m}}(r,z)$ for $m\gg1$. \wmk{The superposition of the four components gives rise to fractional-m WGM-like modes. The CCW traveling WGMs \kss{are} generated due to internal modal coupling with the excited CW modes.} $\bar{m}=(|m_1|+|m_2|)/2$ and $\tilde{m}=(|m_2|-|m_1|)/2$ are the numbers describing the angular momentum and envelope modulation, respectively, of the mixed WGMs. Notably, the complete basis of four mixed modes can be written as $E_{\bar{m}} \cos(\bar{m}\phi \pm \pi/4 + \phi_0)\cos(\tilde{m}\phi \pm \pi/4 + \phi_1)$ when the relative phase of the traveling waves are taken into consideration. The first $\pm\pi/4$ corresponds to the dielectric and air band, and second one represents the two orthogonal envelope distributions. $\phi_0$ describes the phase offset given by the geometry of the photonic crystal modulation, while $\phi_1$ arises from fabrication imperfection. \wmk{\ks{Later,} we will show how to \xl{fix $\phi_1$ beyond this imperfection}.}



\begin{figure}[t!]
\centering\includegraphics[width=0.85\linewidth]{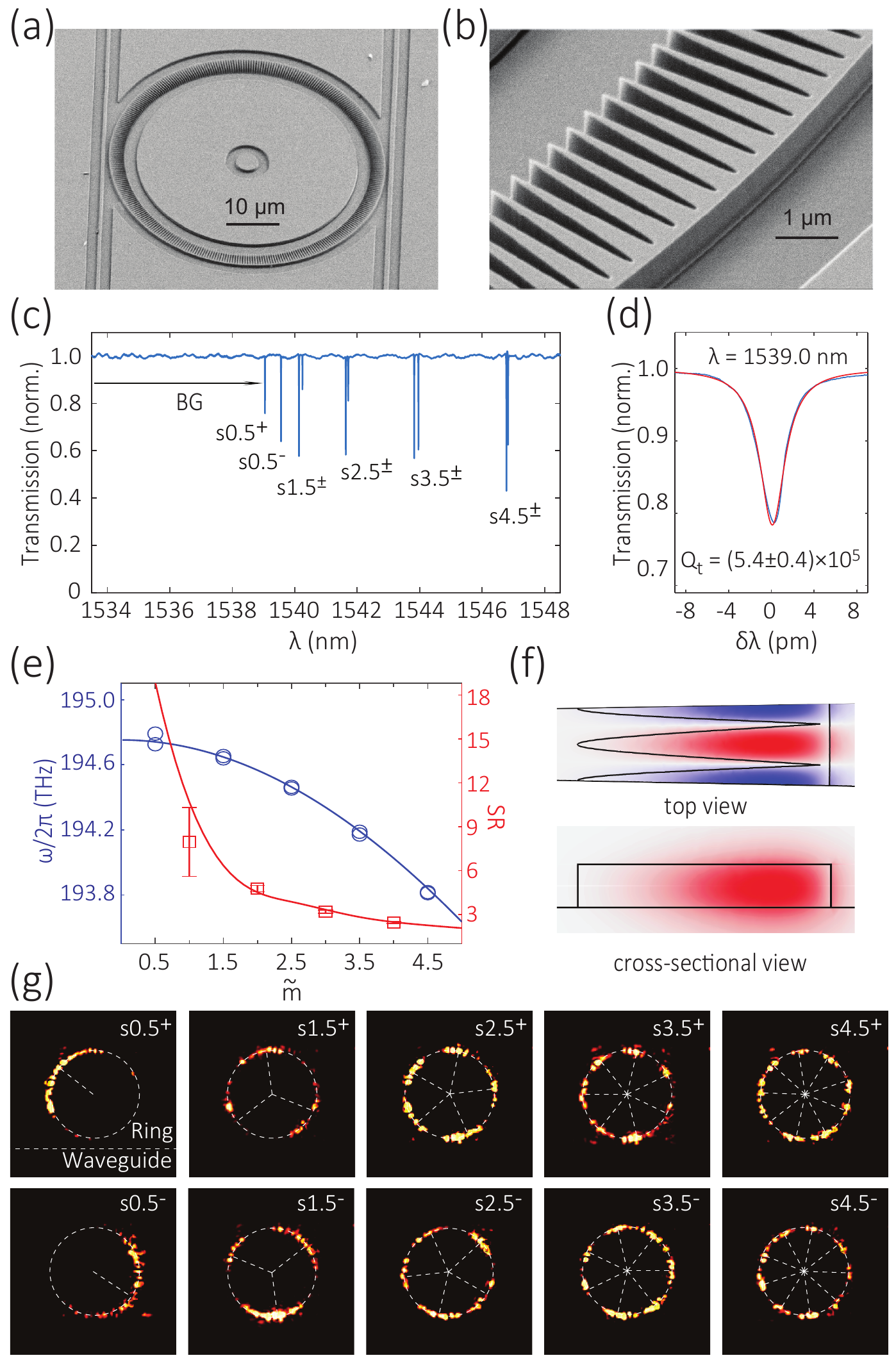}
\caption{\textbf{Fractional angular momentum in a `microgear' photonic crystal ring (MPhCR).} \textbf{a,b,} Scanning electron microscope image of the MPhCR device and zoom-in image of the `microgear' structure. The number of unit cells is $N$=333, resulting in a fractional angular momentum value $\bar{m}$=166.5 for the modes discussed below. \textbf{c,} Linear transmission spectrum of the MPhCR, showing a few pairs of fractional-$m$ slow light modes at the 
dielectric band-edge (labeled $\{s0.5^{\pm},s1.5^{\pm},s2.5^{\pm},s3.5^{\pm},s4.5^{\pm}\}$). These modes results from the hybridization of $m=\{(\pm 166,\pm 167), (\pm 165,\pm 168), (\pm 164,\pm 169), (\pm 163,\pm 170) ,(\pm 162,$ $\pm 171)\}$. \textbf{d,} Zoom-in of the $s0.5^{+}$ mode (blue), along with a nonlinear least squares fit to resonance (red). \textbf{e,} The frequencies (blue) and the slow down ratios (red) of each fractional-$m$ mode near the dielectric band-edge. \textbf{f,} Finite-element method simulated mode profile in one unit cell for the $s0.5^{+}$ mode at 194.25~THz (top and cross-section views). \textbf{g,} Infrared images of the fractional-$m$ modes from a top view of the MPhCR. The dashed lines mark the antinodes (local maxima) of the light intensity distibution.} 
\label{Fig2}
\end{figure}

When $N$ is odd, the hybridization of four modes of $|m_1-m_2| = 1,3,5,...$ gives mixed modes with fractional $\bar{m}=N/2$ and $\tilde{m} = 0.5, 1.5, 2.5,...$, respectively (i.e. the corresponding modes with the same color in Fig. 1(a) and 1(b)). For example, the four WGMs with $|m_1|=5$ and $|m_2|=6$ shown in Fig.~\ref{Fig1}(c) hybridize in a PhCR of $N=11$, generating four mixed WGMs of $\bar{m}=5.5$ and $\tilde{m}=0.5$ shown in Fig. 1(d). Upper (lower) modes correspond to the air (dielectric) band, and left/right are the two orientations of the envelope.

With the fractional angular momentum, phase matching cannot be fulfilled in one round trip of light around the ring. Instead, the mixed WGMs with fractional angular momentum $\bar{m} = N/2$ should undergo two round trips before the phase is matched. Figure~\ref{Fig1}(e) shows the M{\"o}bius-type angular momentum phase and envelope phase topologies of a hybridized mode in the upper and lower panels. For the dots on the edges of the Mobius strip in each panel, their azimuthal angle about the propagation path shows the local phase (i.e. $\bar{m}\phi$ and $\tilde{m}\phi$). After the first round trip (red dots), it has $\pi$ phase delay. The second round trip (blue dots) fulfills the phase matching with an additional $\pi$ phase delay, so that the requisite $2\pi$ phase delay is reached. Their projection on the $x-y$ plane shows the local amplitude $\propto \cos(\bar{m}\phi)$ and $\propto \cos(\tilde{m}\phi)$, respectively. The mode shape of mixed WGMs with $\{\bar{m},\tilde{m}\}=\{5.5,0.5\}$ is shown in the middle panel where $N=11$.

In a real experiment, $\bar{m}$ is a number much larger than that in the toy schematic shown above. For example, we use a photonic crystal ring to demonstrate fractional-$m$ with $\bar{m} = 166.5$. Figure~\ref{Fig2}(a) shows the scanning electron microscope (SEM) images of the nanofabricated MPhCRs made from a 500-nm-thick stoichiometic silicon nitride layer. The ring has a radius of $\approx$~25~$\mu$m and is spatially modulated with a nominal average width $\approx$~1500~nm and modulation amplitude $\approx$~1400~nm, as shown in Fig.~\ref{Fig2}(b). Such a device 
supports fundamental transverse-electric-like (TE) modes, whose dominant electric field is in the radial direction and can be theoretically modeled by Eq.~(\ref{Eq1}). We carry out finite-element method simulations of these devices over a unit cell. Top and cross-sectional views of the generated field profile are shown in Fig.~\ref{Fig2}(f), corresponding to $E_{\bar{m}}(r,z_0) \cos(\bar{m}\phi)$ and $E_{\bar{m}}(r,z)$, respectively, in Eq.~(\ref{Eq1}), where $z=z_0$ corresponds to a plane located midway through the thickness of the ring. The dielectric mode is well confined in the silicon nitride core, preserving conventional WGM profiles. We couple light to the device via on-chip waveguides (Fig.~\ref{Fig2}(a)), and its normalized transmission spectrum is shown in Fig.~\ref{Fig2}(c). The MPhCR with odd $N = 2\times 166.5=333$ opens a bandgap between the $m=166$ and $m=167$ WGMs, hybridizing them and their neighbors, and generating the mixed modes $\{s\tilde{m}^{+},s\tilde{m}^{-},a\tilde{m}^{+},a\tilde{m}^{-}\}$ where $\tilde{m}=0.5, 1.5, 2.5, ...$

\begin{figure*}[t!]
\centering\includegraphics[width=0.80\linewidth]{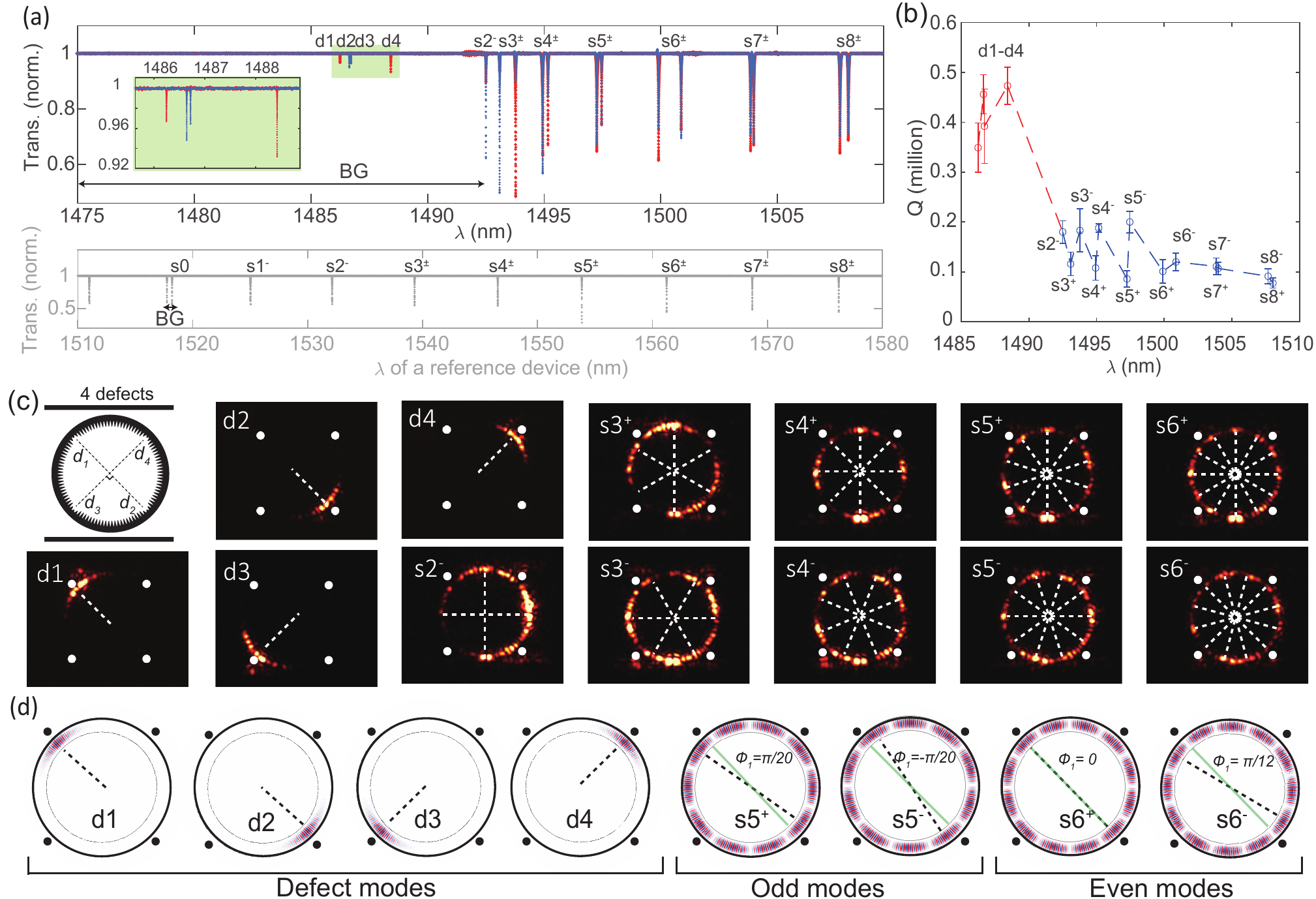}
\caption{\color{black}\textbf{Photonic crystal ring with multiple defects.} \textbf{a,} The normalized transmission spectrum of a photonic crystal microring with four symmetric defects is shown on the top panel. Red (blue) line corresponds to a measurement taken with the \kss{upper (lower)} waveguide shown in (c). The spectrum of a reference device with small modulation ($A=20$ nm) and without defects is illustrated as gray dots on the lower panel. \textbf{b,} Intrinsic quality factor of the modes. \textbf{c,} Schematic of the PhCR with four defects and IR image of the modes up to $\tilde{m}=6$. The white dots label the \ks{angular} position of the symmetric defects ($\pi/2$ relative phase). The white dashed lines highlight the antinodes of the mode. The four defects modes \ks{(labeled d1-d4)} are highly localized. The symmetry of the defects decides the orientation of the WGM-like modes \ks{(labeled s2 to s6)}. \textbf{d,} Schematic of the mode shape. The black dots label the \ks{angular} location of the defects. The black dashed lines denote the orientation of the modes. \ks{For the non-localized WGM-like modes, the} green lines denote the orientation of the defects. The relative angle between the two lines is $\phi_1$.} 
\label{Fig3}
\end{figure*}
The fractional-$m$ WGMs preserve all the main features of conventional WGMs, including high-$Q$, ease in design and fabrication, and \kss{controllable} resonator-waveguide coupling. As shown in Fig.~\ref{Fig2}(d), the total optical $Q$ ($Q_t$) of the mode $s0.5^+$ is measured to be $Q_t=(5.4\pm0.4)\times10^5$, comparable to conventional WGMs in microrings. The mode spectrum is compressed near the band-edge, distinct from that of an unmodulated ring. \kss{In particular}, the reduced FSR indicates a decrease in group velocity. \wmk{The slowdown ratio ($SR$) defines the factor by which the group velocity in the MPhCR is reduced compared to that in the conventional silicon nitride ring\cite{Lee_OL_2012}.} The FSR between s0.5 and s1.5 is measured to be ($122\pm33$) GHz, corresponding to $SR$ = $8.0\pm2.3$ where the uncertainty is arising from the mode splitting of s0.5 and s1.5. We collect the frequencies of the first five pairs of modes in Fig.~\ref{Fig2}(c), and calculate their SRs, as shown in Fig.~\ref{Fig2}(e). The simulated frequencies (blue curve) and SR ratios (red curve) with \kss{the} MIT Photonic Bands method\cite{Johnson_OE_2001} show good agreement with experimental data. Additionally, in the dielectric band, the mode splitting of degenerate modes near the band-edge is much larger than for other modes (Fig.~\ref{Fig2}(c)), indicating stronger coupling near the band-edge. It can be attributed to the slow light enhanced coupling, i.e. higher group index leads to a larger backscattering rate~\cite{morichetti2010roughness}, although its interplay with fabrication imperfection requires further investigation. \wmk{The splitting of modes far from the band-edge is attributed to \xl{fabrication} imperfections.}


Infrared images of scattered light from the hybridized dielectric band-edge modes are illustrated in Fig.~\ref{Fig2}(g). The images in Fig.~\ref{Fig2}(g) clearly display the number of envelope antinodes to be odd for each displayed mode, and equal to 2$\tilde{m}$. Remarkably, different from integer-$m$ WGMs, the fundamental dielectric mode $s0.5^{+}$ and $s0.5^{-}$ are localized orthogonally at opposite sides of the ring. This mode localization can be beneficial for increasing coupling to other degrees of freedom. However, the orientation ($\phi_1$) of these two modes is currently random, dictated by fabrication imperfections. 


\wmk{The orientation of the WGM-like modes can be controlled by breaking the continuous rotational symmetry via introducing artificial defects.} \ks{The mode localization produced by the introduction of a single defect in a device with an even number of modulations was studied in Ref.~\onlinecite{Lu_NatPhoton_2022}. The Supplementary Material presents a similar study for a single defect in a device with an odd number of modulations, where the localized defect state co-exists with band-edge states exhibiting fractional angular momentum. Here, we introduce multiple defects in a device with an even number of modulations.} \wmk{Figure 3 shows a MPhCR with 4 symmetrical defects. Each defect has $\pi/4$ misalignment with the adjacent waveguide as shown in the schematic of Fig. 3(c). Figure 3(a) shows the normalized transmission spectrum of the MPhCR on the top panel. Four modes are localized to the defects, indicated by their wavelength shift \kss{into} the bandgap (BG). The modes are highly localized,} \ks{evident in transmission spectroscopy because they only appear in the spectrum measured by the adjacent waveguide (red/blue line, respectively), and more directly through IR imaging of the device as shown in Fig.~\ref{Fig3}(c).} \wmk{The spectrum of a reference device with small modulation ($A=20$ nm) and without defects is illustrated as gray dots on the lower panel in Fig.~\ref{Fig3}(a). The small modulation only opens a small BG without disturbing other WGMs whose mode splitting is barely visible. Comparing the multi-defect device with the reference, we get the first important conclusion that the mode number is conserved, i.e. four defects localize four modes while other modes \reply{remain} nearly unperturbed.} \ks{The second point} \wmk{is that the orientation of the non-localized modes is controllable by the introduced defects. The introduced defects break the continuous rotational symmetry while creating a new discrete rotational symmetry with a $\pi$/2 period. The renormalized WGMs have period of $\pi/\tilde{m}$. For the modes with even $\tilde{m}$ = 2, 4, 6..., they can match the discrete rotational symmetry of $\pi$/2, so they show orientation $\phi_1 = 0$ and $\phi_1 = \pi/2\tilde{m}$. For the odd modes of $\tilde{m}$ = 1, 3, 5..., instead, they show orientation $\phi_1 = \pm\pi/4\tilde{m}$, as shown in Fig.~3(c) and (d). \reply{Such orientation control based on discrete rotational symmetry also generally applies to devices with other numbers of symmetric defects (see Supplementary Material). Even a single large artificial defect can provide such control.} Figure 3(b) shows the measured intrinsic quality factor of the modes. Surprisingly, the defects modes present higher \kss{$Q$} than the WGM-like modes.} \ks{In particular, the defect mode $Qs$ are as high as $\approx$5$\times10^5$, similar to \ks{that for} single defect modes in Ref.~\onlinecite{Lu_NatPhoton_2022}. However, the WGM-like modes, which exhibit high-$Q$ in cavities without defects (Fig.~\ref{Fig2}) and cavities with a single defect~\cite{Lu_NatPhoton_2022}, now show a reduction in $Q$ by a factor of 5$\times$-10$\times$. \ks{The former} can potentially be attributed to the high-degree of localization by the defects, which lowers the possibility of the modes to encounter random imperfections on the ring. Conversely, the WGM-like modes interact with all four defects, and though this \kss{controls their orientation}, it may also increase their radiation and scattering loss.} \ks{Finally, in Fig. 3(a), we see that as we move away from the bandedge, the even modes generally exhibit larger splittings than the odd modes. This is a function of their orientation with respect to the defects(Fig.~\ref{Fig3}(d)), with the two even modes of a given pair showing complete alignment and misalignment with the defects (nondegenerate), while the two odd modes are in principle equally well misaligned (degenerate).}


In summary, \wmk{we have demonstrated} \wmk{understanding of renormalization and localization of WGMs near the bandedge of a photonic crystal ring.} Based on it, we presented optical modes with fractional (half-integer) angular momentum. The photonic crystal patterning enables controlled hybridization of WGMs into half-integer angular momentum states that retain high-$Q$ while exhibiting reduced group velocity. Furthermore, \wmk{we demonstrate a multi-mode localization process} \ks{based on the introduction of multiple \reply{symmetric} artificial defects. These defects strongly localize modes that are closest to the bandedge, while impacting the orientation of modes farther away from the bandedge, which largely retain their WGM-like character. These far modes' orientation and splittings are controlled by the defect-induced continuous rotational symmetry breaking.} \reply{The discrete rotational symmetry of the defects provides an intuitive understanding to realize the mode orientation and degeneracy control. Modal-defect interaction without this symmetry is another interesting topic, requiring further investigation.} We anticipate that our platform may broaden the usage of fractional-$m$ light in applications including nonlinear photonics~\cite{Moss_NatPhoton_2013}, quantum photonics~\cite{Obrien_NatPhoton_2009}, and cavity optomechanics~\cite{Aspelmeyer_RevModPhys_2014}.

\medskip
See Supplementary Materials for further discussions on fractional optical angular mometnum generation, which include Refs.~[\onlinecite{Wilczek_PRL_1982, wilczek1982magnetic, Molchan_JOA_2008, Jesus-Silva_OL_2012, Martinez-Castellanos_JOSAA_2013, Martinez-Castellanos_OL_2015, Wang_SR_2016, Pisanty_NatPhoton_2019}].



\medskip
This work is supported by the DARPA SAVaNT and NIST-on-a-chip programs, and partly sponsored by the Army Research Office under Cooperative Agreement Number W911NF-21-2-0106. M.W. is supported by the cooperative research agreement between University of Maryland and NIST, Award no. 70NANB10H193. The authors thank Kaikai Liu for helpful discussions.



\vspace{30pt} 
\begin{center} 
\textbf{Supplementary Information}
\end{center}

\section{Experimental Setup}
 \ks{A schematic of the experimental setup is shown in Fig.~\ref{FigS1}, where a broadly tunable, fiber-coupled external cavity diode laser (wavelength range from 1470~nm to 1640~nm) is coupled into a photonic chip via \xl{a} lensed optical fiber and the output signal is out-coupled via a lensed optical fiber \xl{to a} photodetector. A data acquisition system displays the swept-wavelength spectrum.  An infrared camera at the output of a long working distance imaging system is used to image the photonic chip, and the scattered light patterns associated with different modes are acquired when the laser \xl{is tuned} into resonance with the modes.}  

\begin{figure}[t!]
\setcounter{figure}{0}
\renewcommand{\figurename}{Fig.}
\renewcommand{\thefigure}{S\arabic{figure}}
\centering\includegraphics[width=\linewidth]{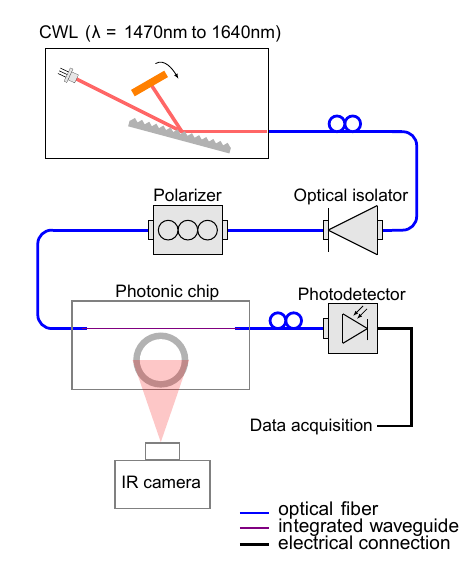}
\caption{\textbf{Experimental setup}. Experimental setup for swept wavelength spectroscopy and imaging of the MPhCR modes. CWL\xl{:} continuous wavelength tunable laser. \xl{IR: infrared.}} 
\label{FigS1}
\end{figure}

\section{Fractional Optical Angular Momentum Generation}
\noindent Spin angular momentum and orbital angular momentum of light, defining the electromagnetic wave oscillation direction (i.e., polarization) and the spatial distribution of field phase, are two quantization numbers describing light's total angular momentum. In analogy to fractional spin particles in two-dimensional systems~\cite{Wilczek_PRL_1982, wilczek1982magnetic}, recent works show that light beams can carry fractional orbital angular momentum~\cite{Gotte_OE_2008, Molchan_JOA_2008, Jesus-Silva_OL_2012, Nugrowati_OE_2012, Martinez-Castellanos_JOSAA_2013, Martinez-Castellanos_OL_2015, Ballantine_SciAdv_2016, Wang_SR_2016, Pisanty_NatPhoton_2019}. A fractional 2$\pi$-phase shift along the azimuth of light’s propagation direction is the signature feature of these works. This unusual quantization can be realized by symmetry breaking via various different mechanisms, such as spiral phase modulation~\cite{Beijersbergen_OptCommun_1994}, biaxial crystals~\cite{Ballantine_SciAdv_2016, Turpin_SR_2017}, astigmatic elements~\cite{Nugrowati_OE_2012,Nugrowati_OC_2013}, differential operators acting on a Gaussian beam~\cite{Gutierrez-Vega_JOA_2008}, and superposition of light modes of different integer angular quantization~\cite{Molina-Terriza_PRL_2002, Gotte_OE_2008}. The above experiments have been done on optical beams in free-space; our work is distinguished by presenting optical modes with fractional orbital angular momentum (fractional-$m$) and properties like conventional WGMs (e.g., high-$Q$ with the same radial profile) within a controllable, chip-integrated platform.

\section{Defect mode localization and fractional-$m$ optical angular momentum}
\begin{figure*}[t!]
\centering\includegraphics[width=0.80\linewidth]{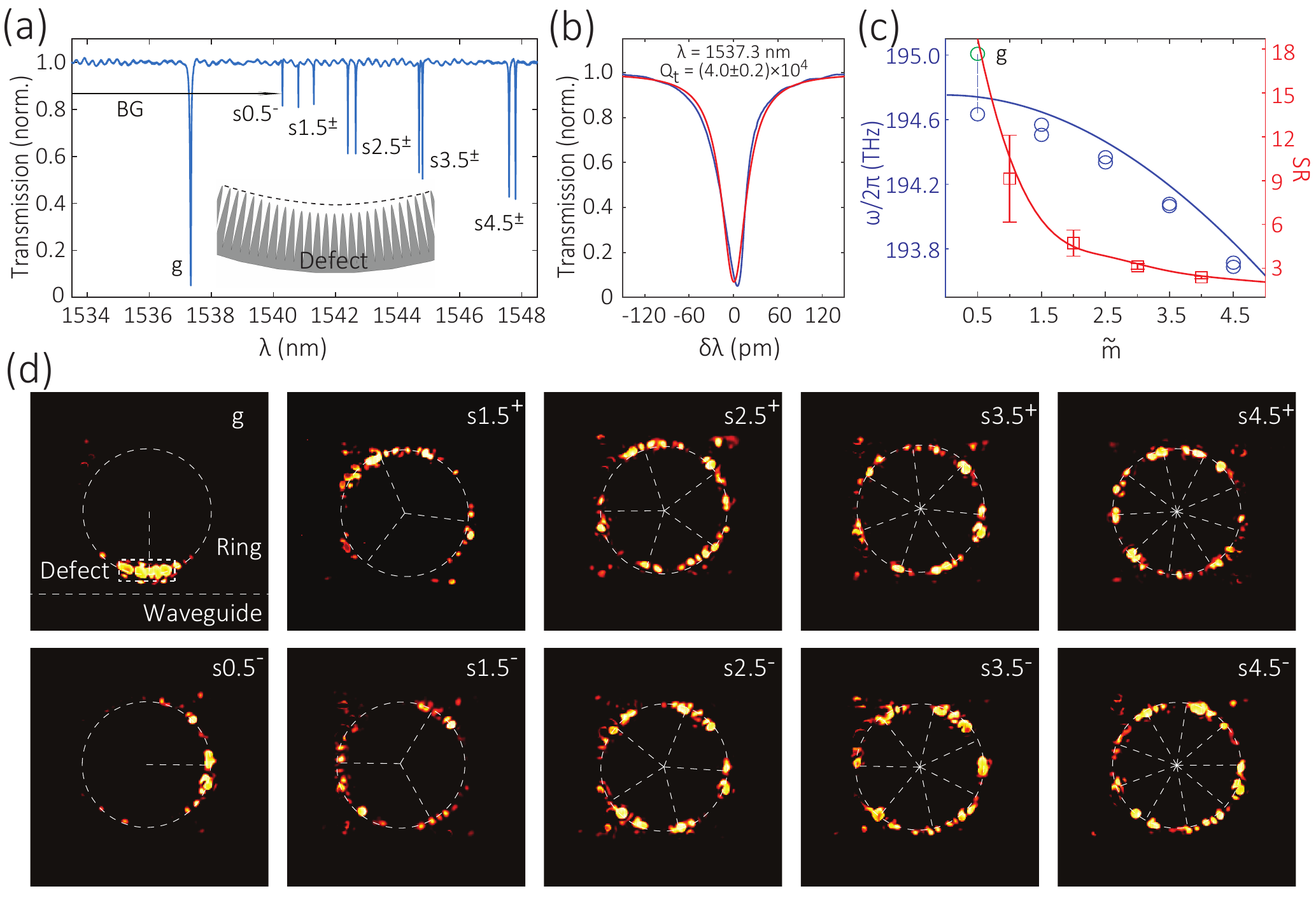}
\renewcommand{\figurename}{Fig.}
\renewcommand{\thefigure}{S\arabic{figure}}
\caption{\textbf{Fractional-$m$ optical angular momentum in defect MPhCRs.} \textbf{a,} Transmission spectrum of the fractional-$m$ defect MPhCR, \ks{showing} the  defect mode (labeled $g$) that is localized from the bandedge slow light mode ($s0.5^+$), and the remaining slow light modes \ks{(all of which exhibit half-integer angular momentum)}. \textbf{b,} Zoom-in of the $g$ mode (blue) and a nonlinear least squares fit to \ks{its transmission resonance} (red). \textbf{c,} The frequency (blue) and $SR$ (red) of each fractional-$m$ mode near the dielectric band-edge. The uncertainty in $SR$ comes from estimating free spectral ranges with the split modes. The uncertainty in the eigenfrequency is from the nonlinear fit. \textbf{d,} Infrared images of the defect mode and the slow light modes of the MPhCR. The dashed lines mark the antinodes (local maxima) of the light intensity \kss{distribution}. The stray light off the rings is arising from reflection at the waveguide-ring junctions.} 
\label{FigS2}
\end{figure*}
\noindent \ks{Figure~\ref{FigS2} presents data on mode localization by a single defect in a half-$m$ MPhCR. We construct a defect region by varying the PhC modulation amplitude quadratically across 24 cells, with a maximum modulation depth deviation of 10~$\%$, as shown in the inset of Figure~\ref{FigS2}(a). Figure~\ref{FigS2}(a) shows the normalized transmission spectrum of the defect MPhCR, where most of \kss{the} modes stay unperturbed relative to \kss{the} non-defect MPhCR, and retain comparable $Q$ (e.g. $s1.5^+$ has $Q_t = (2.1\pm0.3)\times10^5$ ). Only one of the fundamental dielectric modes is pulled into the bandgap, becoming a defect mode ($g$), with $Q_t = (4.0\pm0.2)\times10^4$, as shown in Fig.~\ref{FigS2}(b). Compared to Fig.~2(e) in the main text, the relatively low total quality factor $Q_t$ is attributed to the much stronger coupling loss to the waveguide adjacent to the defect. Given that this is a standing wave resonance in which strong overcoupling corresponds to a transmission level near zero, it is difficult to accurately extract the intrinsic quality factor from this data. The mechanism of mode selection in localization is still under investigation, though we note that in this case, it was the mode closest to the band-edge ($s0.5^{+}$) from which $g$ is localized. Figure~\ref{FigS2}(c) displays the dielectric band-edge mode frequencies (blue) and $SR$ (red) for the modes shown in Fig.~\ref{FigS2}(a). The $SR$ of the modes near the band-edge is measured to be as large as 9.1$\pm$3.0, indicating that the slow light property of fractional-$m$ WGMs is not degraded by the introduced defect. Infrared images of the defect mode and fractional-$m$ WGMs are shown in Fig.~\ref{FigS2}(d). Notably, the $g$ mode of fractional-$m$ WGMs is highly localized at the position of the defect at the bottom adjacent to the waveguide. The different depths of the transmission dips in Fig.~\ref{FigS2}(a) are attributed to their different coupling rates to the acess waveguide in the under-coupled regime. For example, the $s0.5^{-}$ and $g$ modes are expected to have the best and worst coupling to the waveguide, respectively, from the infared images in Fig.~\ref{FigS2}(d), and indeed they exhibit the deepest and shallowest transmission dips, respectively.} 

\section{Mode orientation results from devices with a differing number of symmetric defects}

\begin{figure*}[t!]
\centering\includegraphics[width=0.80\linewidth]{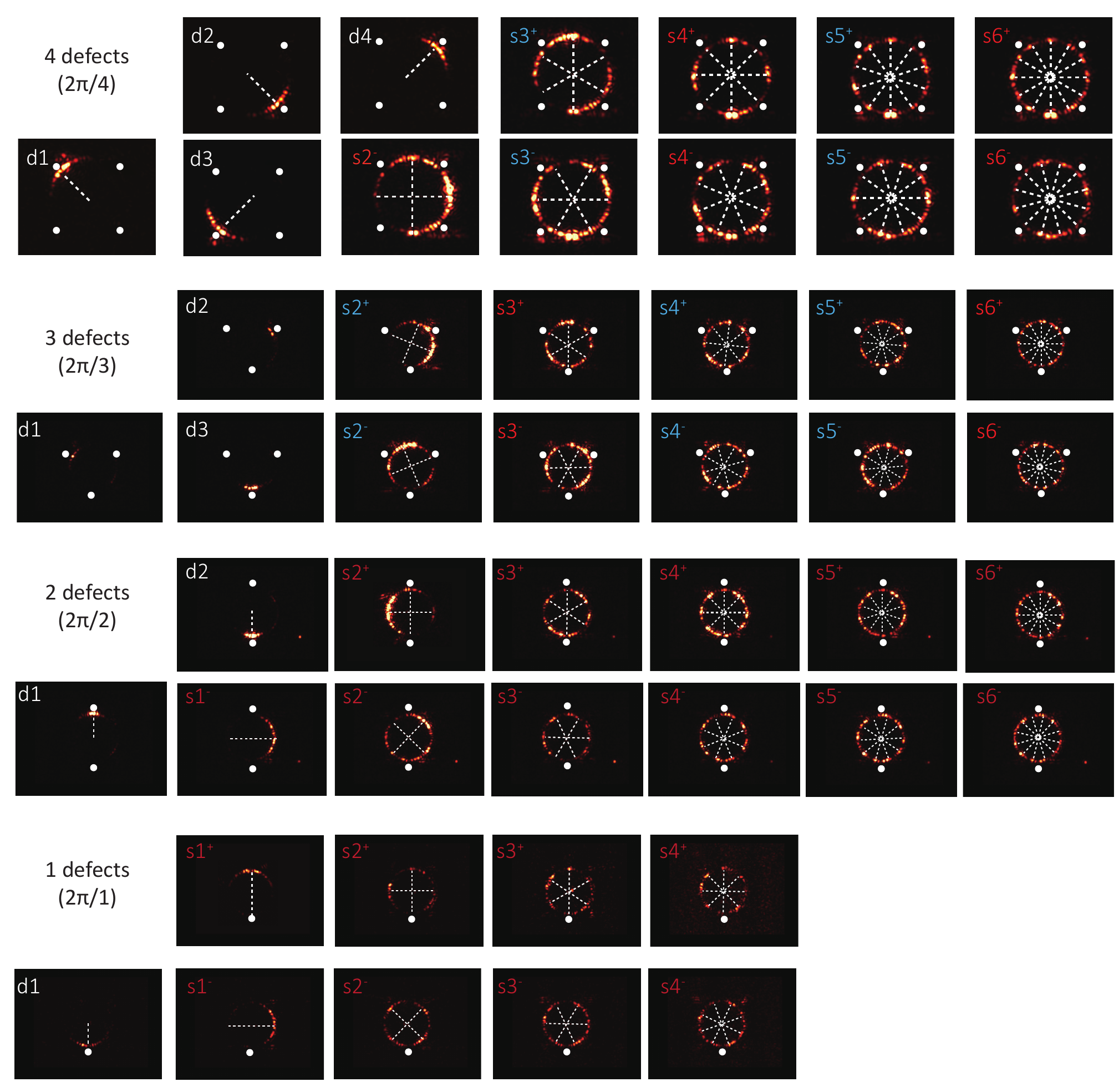}
\renewcommand{\figurename}{Fig.}
\renewcommand{\thefigure}{S\arabic{figure}}
\caption{\kss{Multi-defect photonic crystal ring with number of defects} $K =$ 1 to 4. The white dots label the location of the defects and the white dashed lines denote the anti-nodes. For a device with $K$ defects, its discrete rotational symmetry period is $2\pi⁄K$ and its first $K$ modes are localized to the defects. The non-localized modes’ orientation is defined by the defects. Red labels modes that match the discrete rotational symmetry period of $2\pi⁄K$, showing the non-degenerate orientation of $\phi_1=0$ or $\pi/2\tilde{m}$. Blue labels modes that do not match the defects’ topology, showing  degenerate orientations of $\phi_1=\pm\pi/4\tilde{m}$.} 
\label{FigS3}
\end{figure*}

\begin{figure*}
  \centering\includegraphics[width=0.80\linewidth]{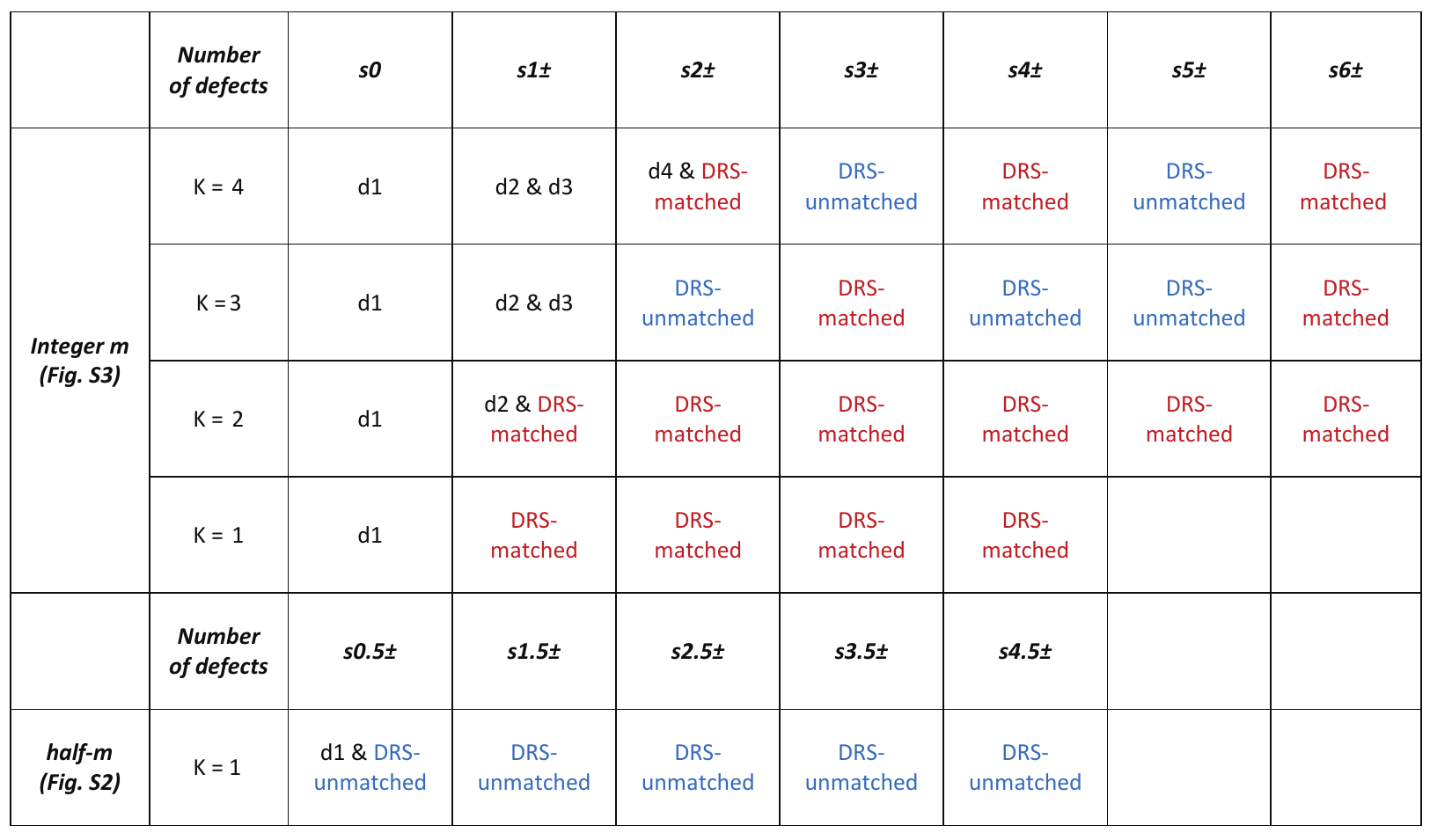}
    \renewcommand{\figurename}{Table}
    \renewcommand{\thefigure}{S1}
    \caption{Summary of modes in a photonic crystal microring with multiple defects. The non-localized modes are the band-edge states $s\tilde{m}^{\pm}$, where $\tilde{m}$ is an integer or half-integer depending on the number of photonic crystal periods, while the localized modes are labeled $dn$, where $n$ is an integer (the total number of localized modes matches the total number of defects). Red color labels non-localized modes that match the discrete rotational symmetry (DRS) of the defects. Blue color labels non-localized modes that do not match the DRS of the defects. All of the modes’ patterns are only decided and can be predicted by an intuitive rule based on DRS, with non-degenerate modes exhibiting $\phi_1=0$ or $\pi/2\tilde{m}$ (red) and degenerate modes exhibiting $\phi_1=\pm\pi/4\tilde{m}$ (blue), respectively.} 
\end{figure*}

\reply{The results shown in Fig. S3 confirm that the defects localize light to small volumes while otherwise maintaining the conventional properties of WGMs. The number of modes (labeled $d1$, $d2$, etc, in Fig. S3) is conserved for defect numbers of 1-4.}  

\reply{The orientation of non-localized modes ($s\tilde{m}\pm$) can be controlled by a single defect or multiple defects. Remarkably, the orientation is only decided by the discrete rotational symmetry of the symmetric defect(s) no matter what \kss{the number of defects} $K$ is. Figure S3 and Table S1 show \kss{experimental results from} four cases with $K =$ 1 to 4. The discrete rotational symmetry period (in terms of radian angle) is given by $2\pi⁄K$.}

\reply{For $K =$ 1 and 2, all the non-localized modes can match the discrete rotational symmetry of 2$\pi$ and $\pi$, respectively, making the modes show the orientation of $\phi_1=0$ or $\pi/2\tilde{m}$. For $K$ = 4, as discussed in the main text, only modes with $\tilde{m}=$2,4,6… (even numbers) can match the discrete rotational symmetry period of $\pi/2$, showing $\phi_1=0$ or $\pi/2\tilde{m}$ while odd modes show orientations of $\phi_1=\pm\pi/4\tilde{m}$. For $K$ = 3, similarly, only modes with $\tilde{m}$=3,6,... can match the discrete rotational symmetry period of $2\pi/3$, showing $\phi_1=0$ or $\pi/2\tilde{m}$, while other modes show orientations of $\phi_1=\pm\pi/4\tilde{m}$.}

\reply{This intuitive rule from topology symmetry applies to all cases studied here. For modes matching $2\pi/K$ period, the two modes are non-degenerate with $\phi_1=0$ or $\pi/2\tilde{m}$. For modes that don’t match the 2$\pi⁄K$ period, the two modes are degenerate with $\phi_1=\pm\pi/4\tilde{m}$.}

\reply{For integer-$m$ devices with a single defect, i.e. the lowest panel of Fig.~S3, all non-localized modes ($s$1-$s$4) match the discrete rotational symmetry of the single defect (2$\pi$), so the modes are with the orientation of $\phi_1=0$ or $\pi/2\tilde{m}$. Remarkably, for half-$m$ devices with a single defect, i.e. Fig.~S2, it becomes the opposite. All the non-localized modes do not match the 2$\pi$ discrete rotational symmetry, therefore, they are with orientation $\phi_1=\pm\pi/4\tilde{m}$. This comparison between half-$m$ and integer-$m$ devices with a single defect clearly shows the validity of the observation.}

\bibliography{Half_m.bib}

\begin{thebibliography}{35}
\expandafter\ifx\csname natexlab\endcsname\relax\def\natexlab#1{#1}\fi
\expandafter\ifx\csname bibnamefont\endcsname\relax
  \def\bibnamefont#1{#1}\fi
\expandafter\ifx\csname bibfnamefont\endcsname\relax
  \def\bibfnamefont#1{#1}\fi
\expandafter\ifx\csname citenamefont\endcsname\relax
  \def\citenamefont#1{#1}\fi
\expandafter\ifx\csname url\endcsname\relax
  \def\url#1{\texttt{#1}}\fi
\expandafter\ifx\csname urlprefix\endcsname\relax\def\urlprefix{URL }\fi
\providecommand{\bibinfo}[2]{#2}
\providecommand{\eprint}[2][]{\url{#2}}

\bibitem[{\citenamefont{Aspelmeyer et~al.}(2014)\citenamefont{Aspelmeyer,
  Kippenberg, and Marquardt}}]{Aspelmeyer_RevModPhys_2014}
\bibinfo{author}{\bibfnamefont{M.}~\bibnamefont{Aspelmeyer}},
  \bibinfo{author}{\bibfnamefont{T.~J.} \bibnamefont{Kippenberg}},
  \bibnamefont{and}
  \bibinfo{author}{\bibfnamefont{F.}~\bibnamefont{Marquardt}},
  \bibinfo{journal}{Rev. Mod. Phys.} \textbf{\bibinfo{volume}{86}},
  \bibinfo{pages}{1391} (\bibinfo{year}{2014}).

\bibitem[{\citenamefont{Shen et~al.}(2020)\citenamefont{Shen, Chang, Liu, Wang,
  Yang, Xiang, Wang, He, Liu, Xie et~al.}}]{shen2020integrated}
\bibinfo{author}{\bibfnamefont{B.}~\bibnamefont{Shen}},
  \bibinfo{author}{\bibfnamefont{L.}~\bibnamefont{Chang}},
  \bibinfo{author}{\bibfnamefont{J.}~\bibnamefont{Liu}},
  \bibinfo{author}{\bibfnamefont{H.}~\bibnamefont{Wang}},
  \bibinfo{author}{\bibfnamefont{Q.-F.} \bibnamefont{Yang}},
  \bibinfo{author}{\bibfnamefont{C.}~\bibnamefont{Xiang}},
  \bibinfo{author}{\bibfnamefont{R.~N.} \bibnamefont{Wang}},
  \bibinfo{author}{\bibfnamefont{J.}~\bibnamefont{He}},
  \bibinfo{author}{\bibfnamefont{T.}~\bibnamefont{Liu}},
  \bibinfo{author}{\bibfnamefont{W.}~\bibnamefont{Xie}}, \bibnamefont{et~al.},
  \bibinfo{journal}{Nature} \textbf{\bibinfo{volume}{582}},
  \bibinfo{pages}{365} (\bibinfo{year}{2020}).

\bibitem[{\citenamefont{Rakovich and Donegan}(2010)}]{rakovich2010photonic}
\bibinfo{author}{\bibfnamefont{Y.~P.} \bibnamefont{Rakovich}} \bibnamefont{and}
  \bibinfo{author}{\bibfnamefont{J.~F.} \bibnamefont{Donegan}},
  \bibinfo{journal}{Laser Photon. Rev.} \textbf{\bibinfo{volume}{4}},
  \bibinfo{pages}{179} (\bibinfo{year}{2010}).

\bibitem[{\citenamefont{Roll et~al.}(1998)\citenamefont{Roll, Kaiser, Lange,
  and Schweiger}}]{roll1998ray}
\bibinfo{author}{\bibfnamefont{G.}~\bibnamefont{Roll}},
  \bibinfo{author}{\bibfnamefont{T.}~\bibnamefont{Kaiser}},
  \bibinfo{author}{\bibfnamefont{S.}~\bibnamefont{Lange}}, \bibnamefont{and}
  \bibinfo{author}{\bibfnamefont{G.}~\bibnamefont{Schweiger}},
  \bibinfo{journal}{J. Opt. Soc. Am. A} \textbf{\bibinfo{volume}{15}},
  \bibinfo{pages}{2879} (\bibinfo{year}{1998}).

\bibitem[{\citenamefont{Chang et~al.}(2017)\citenamefont{Chang, Timmermans, and
  Otto}}]{Chang_OE_2017}
\bibinfo{author}{\bibfnamefont{L.}~\bibnamefont{Chang}},
  \bibinfo{author}{\bibfnamefont{F.}~\bibnamefont{Timmermans}},
  \bibnamefont{and} \bibinfo{author}{\bibfnamefont{C.}~\bibnamefont{Otto}},
  \bibinfo{journal}{Opt. Express} \textbf{\bibinfo{volume}{25}},
  \bibinfo{pages}{28946} (\bibinfo{year}{2017}).

\bibitem[{\citenamefont{Wang et~al.}(2021)\citenamefont{Wang, Liu, Liu, Xiao,
  Wang, Fan, Han, Ge, and Song}}]{Wang_LSA_2021}
\bibinfo{author}{\bibfnamefont{S.}~\bibnamefont{Wang}},
  \bibinfo{author}{\bibfnamefont{S.}~\bibnamefont{Liu}},
  \bibinfo{author}{\bibfnamefont{Y.}~\bibnamefont{Liu}},
  \bibinfo{author}{\bibfnamefont{S.}~\bibnamefont{Xiao}},
  \bibinfo{author}{\bibfnamefont{Z.}~\bibnamefont{Wang}},
  \bibinfo{author}{\bibfnamefont{Y.}~\bibnamefont{Fan}},
  \bibinfo{author}{\bibfnamefont{J.}~\bibnamefont{Han}},
  \bibinfo{author}{\bibfnamefont{L.}~\bibnamefont{Ge}}, \bibnamefont{and}
  \bibinfo{author}{\bibfnamefont{Q.}~\bibnamefont{Song}},
  \bibinfo{journal}{Light Sci. Appl.} \textbf{\bibinfo{volume}{10}},
  \bibinfo{pages}{135} (\bibinfo{year}{2021}).

\bibitem[{\citenamefont{Beijersbergen et~al.}(1994)\citenamefont{Beijersbergen,
  Coerwinkel, Kristensen, and Woerdman}}]{Beijersbergen_OptCommun_1994}
\bibinfo{author}{\bibfnamefont{M.~W.} \bibnamefont{Beijersbergen}},
  \bibinfo{author}{\bibfnamefont{R.~P.} \bibnamefont{Coerwinkel}},
  \bibinfo{author}{\bibfnamefont{M.}~\bibnamefont{Kristensen}},
  \bibnamefont{and} \bibinfo{author}{\bibfnamefont{J.~P.}
  \bibnamefont{Woerdman}}, \bibinfo{journal}{Opt. Commun.}
  \textbf{\bibinfo{volume}{112}}, \bibinfo{pages}{179} (\bibinfo{year}{1994}).

\bibitem[{\citenamefont{Molina-Terriza
  et~al.}(2002)\citenamefont{Molina-Terriza, Torres, and
  Torner}}]{Molina-Terriza_PRL_2002}
\bibinfo{author}{\bibfnamefont{G.}~\bibnamefont{Molina-Terriza}},
  \bibinfo{author}{\bibfnamefont{J.~P.} \bibnamefont{Torres}},
  \bibnamefont{and} \bibinfo{author}{\bibfnamefont{L.}~\bibnamefont{Torner}},
  \bibinfo{journal}{Phys. Rev. Lett.} \textbf{\bibinfo{volume}{88}},
  \bibinfo{pages}{013601} (\bibinfo{year}{2002}).

\bibitem[{\citenamefont{G{\"{o}}tte et~al.}(2008)\citenamefont{G{\"{o}}tte,
  O'Holleran, Preece, Flossmann, Franke-Arnold, Barnett, and
  Padgett}}]{Gotte_OE_2008}
\bibinfo{author}{\bibfnamefont{J.~B.} \bibnamefont{G{\"{o}}tte}},
  \bibinfo{author}{\bibfnamefont{K.}~\bibnamefont{O'Holleran}},
  \bibinfo{author}{\bibfnamefont{D.}~\bibnamefont{Preece}},
  \bibinfo{author}{\bibfnamefont{F.}~\bibnamefont{Flossmann}},
  \bibinfo{author}{\bibfnamefont{S.}~\bibnamefont{Franke-Arnold}},
  \bibinfo{author}{\bibfnamefont{S.~M.} \bibnamefont{Barnett}},
  \bibnamefont{and} \bibinfo{author}{\bibfnamefont{M.~J.}
  \bibnamefont{Padgett}}, \bibinfo{journal}{Opt. Express}
  \textbf{\bibinfo{volume}{16}}, \bibinfo{pages}{993} (\bibinfo{year}{2008}).

\bibitem[{\citenamefont{Guti{\'{e}}rrez-Vega and
  L{\'{o}}pez-Mariscal}(2008)}]{Gutierrez-Vega_JOA_2008}
\bibinfo{author}{\bibfnamefont{J.~C.} \bibnamefont{Guti{\'{e}}rrez-Vega}}
  \bibnamefont{and}
  \bibinfo{author}{\bibfnamefont{C.}~\bibnamefont{L{\'{o}}pez-Mariscal}},
  \bibinfo{journal}{J. Opt. A: Pure Appl. Opt.} \textbf{\bibinfo{volume}{10}},
  \bibinfo{pages}{015009} (\bibinfo{year}{2008}).

\bibitem[{\citenamefont{Nugrowati et~al.}(2012)\citenamefont{Nugrowati, Stam,
  and Woerdman}}]{Nugrowati_OE_2012}
\bibinfo{author}{\bibfnamefont{A.~M.} \bibnamefont{Nugrowati}},
  \bibinfo{author}{\bibfnamefont{W.~G.} \bibnamefont{Stam}}, \bibnamefont{and}
  \bibinfo{author}{\bibfnamefont{J.~P.} \bibnamefont{Woerdman}},
  \bibinfo{journal}{Opt. Express} \textbf{\bibinfo{volume}{20}},
  \bibinfo{pages}{27429} (\bibinfo{year}{2012}).

\bibitem[{\citenamefont{Nugrowati and Woerdman}(2013)}]{Nugrowati_OC_2013}
\bibinfo{author}{\bibfnamefont{A.~M.} \bibnamefont{Nugrowati}}
  \bibnamefont{and} \bibinfo{author}{\bibfnamefont{J.~P.}
  \bibnamefont{Woerdman}}, \bibinfo{journal}{Opt. Commun.}
  \textbf{\bibinfo{volume}{308}}, \bibinfo{pages}{253} (\bibinfo{year}{2013}).

\bibitem[{\citenamefont{Ballantine et~al.}(2016)\citenamefont{Ballantine,
  Donegan, and Eastham}}]{Ballantine_SciAdv_2016}
\bibinfo{author}{\bibfnamefont{K.~E.} \bibnamefont{Ballantine}},
  \bibinfo{author}{\bibfnamefont{J.~F.} \bibnamefont{Donegan}},
  \bibnamefont{and} \bibinfo{author}{\bibfnamefont{P.~R.}
  \bibnamefont{Eastham}}, \bibinfo{journal}{Sci. Adv.}
  \textbf{\bibinfo{volume}{2}}, \bibinfo{pages}{1501748}
  (\bibinfo{year}{2016}).

\bibitem[{\citenamefont{Turpin et~al.}(2017)\citenamefont{Turpin, Rego,
  Pic\'{o}n, Rom\'{a}n, and Hern\'{a}ndez-Garc\'{i}a}}]{Turpin_SR_2017}
\bibinfo{author}{\bibfnamefont{A.}~\bibnamefont{Turpin}},
  \bibinfo{author}{\bibfnamefont{L.}~\bibnamefont{Rego}},
  \bibinfo{author}{\bibfnamefont{A.}~\bibnamefont{Pic\'{o}n}},
  \bibinfo{author}{\bibfnamefont{J.~S.} \bibnamefont{Rom\'{a}n}},
  \bibnamefont{and}
  \bibinfo{author}{\bibfnamefont{C.}~\bibnamefont{Hern\'{a}ndez-Garc\'{i}a}},
  \bibinfo{journal}{Sci. Rep.} \textbf{\bibinfo{volume}{7}},
  \bibinfo{pages}{43888} (\bibinfo{year}{2017}).

\bibitem[{\citenamefont{Peng et~al.}(2016)\citenamefont{Peng, {\"O}zdemir,
  Liertzer, Chen, Kramer, Y{\i}lmaz, Wiersig, Rotter, and
  Yang}}]{peng2016chiral}
\bibinfo{author}{\bibfnamefont{B.}~\bibnamefont{Peng}},
  \bibinfo{author}{\bibfnamefont{{\c{S}}.~K.} \bibnamefont{{\"O}zdemir}},
  \bibinfo{author}{\bibfnamefont{M.}~\bibnamefont{Liertzer}},
  \bibinfo{author}{\bibfnamefont{W.}~\bibnamefont{Chen}},
  \bibinfo{author}{\bibfnamefont{J.}~\bibnamefont{Kramer}},
  \bibinfo{author}{\bibfnamefont{H.}~\bibnamefont{Y{\i}lmaz}},
  \bibinfo{author}{\bibfnamefont{J.}~\bibnamefont{Wiersig}},
  \bibinfo{author}{\bibfnamefont{S.}~\bibnamefont{Rotter}}, \bibnamefont{and}
  \bibinfo{author}{\bibfnamefont{L.}~\bibnamefont{Yang}},
  \bibinfo{journal}{Proc. Natl. Acad. Sci.} \textbf{\bibinfo{volume}{113}},
  \bibinfo{pages}{6845} (\bibinfo{year}{2016}).

\bibitem[{\citenamefont{Wang et~al.}(2020)\citenamefont{Wang, Jiang, Zhao,
  Zhang, Hsu, Peng, Stone, Jiang, and Yang}}]{wang2020electromagnetically}
\bibinfo{author}{\bibfnamefont{C.}~\bibnamefont{Wang}},
  \bibinfo{author}{\bibfnamefont{X.}~\bibnamefont{Jiang}},
  \bibinfo{author}{\bibfnamefont{G.}~\bibnamefont{Zhao}},
  \bibinfo{author}{\bibfnamefont{M.}~\bibnamefont{Zhang}},
  \bibinfo{author}{\bibfnamefont{C.~W.} \bibnamefont{Hsu}},
  \bibinfo{author}{\bibfnamefont{B.}~\bibnamefont{Peng}},
  \bibinfo{author}{\bibfnamefont{A.~D.} \bibnamefont{Stone}},
  \bibinfo{author}{\bibfnamefont{L.}~\bibnamefont{Jiang}}, \bibnamefont{and}
  \bibinfo{author}{\bibfnamefont{L.}~\bibnamefont{Yang}},
  \bibinfo{journal}{Nat. Phys.} \textbf{\bibinfo{volume}{16}},
  \bibinfo{pages}{334} (\bibinfo{year}{2020}).

\bibitem[{\citenamefont{Xu et~al.}(2019)\citenamefont{Xu, Shi, Guo, Dong, and
  Zou}}]{Xu_APL_2019}
\bibinfo{author}{\bibfnamefont{X.-B.} \bibnamefont{Xu}},
  \bibinfo{author}{\bibfnamefont{L.}~\bibnamefont{Shi}},
  \bibinfo{author}{\bibfnamefont{G.-C.} \bibnamefont{Guo}},
  \bibinfo{author}{\bibfnamefont{C.-H.} \bibnamefont{Dong}}, \bibnamefont{and}
  \bibinfo{author}{\bibfnamefont{C.-L.} \bibnamefont{Zou}},
  \bibinfo{journal}{Appl. Phys. Lett.} \textbf{\bibinfo{volume}{114}},
  \bibinfo{pages}{101106} (\bibinfo{year}{2019}).

\bibitem[{\citenamefont{Lu et~al.}(2022)\citenamefont{Lu, McClung, and
  Srinivasan}}]{Lu_NatPhoton_2022}
\bibinfo{author}{\bibfnamefont{X.}~\bibnamefont{Lu}},
  \bibinfo{author}{\bibfnamefont{A.}~\bibnamefont{McClung}}, \bibnamefont{and}
  \bibinfo{author}{\bibfnamefont{K.}~\bibnamefont{Srinivasan}},
  \bibinfo{journal}{Nat. Photon.} \textbf{\bibinfo{volume}{16}},
  \bibinfo{pages}{66} (\bibinfo{year}{2022}).

\bibitem[{\citenamefont{Lee and Fauchet}(2012)}]{Lee_OL_2012}
\bibinfo{author}{\bibfnamefont{J.~Y.} \bibnamefont{Lee}} \bibnamefont{and}
  \bibinfo{author}{\bibfnamefont{P.~M.} \bibnamefont{Fauchet}},
  \bibinfo{journal}{Opt. Lett.} \textbf{\bibinfo{volume}{37}},
  \bibinfo{pages}{58} (\bibinfo{year}{2012}).

\bibitem[{\citenamefont{Zhang et~al.}(2015)\citenamefont{Zhang, Qiu, Zeng, Li,
  Gao, Wang, Yu, and Xia}}]{Zhang_PTL_2015}
\bibinfo{author}{\bibfnamefont{Y.}~\bibnamefont{Zhang}},
  \bibinfo{author}{\bibfnamefont{X.}~\bibnamefont{Qiu}},
  \bibinfo{author}{\bibfnamefont{C.}~\bibnamefont{Zeng}},
  \bibinfo{author}{\bibfnamefont{D.}~\bibnamefont{Li}},
  \bibinfo{author}{\bibfnamefont{G.}~\bibnamefont{Gao}},
  \bibinfo{author}{\bibfnamefont{Y.}~\bibnamefont{Wang}},
  \bibinfo{author}{\bibfnamefont{J.}~\bibnamefont{Yu}}, \bibnamefont{and}
  \bibinfo{author}{\bibfnamefont{J.}~\bibnamefont{Xia}}, \bibinfo{journal}{IEEE
  Photon. Tech. Lett.} \textbf{\bibinfo{volume}{27}}, \bibinfo{pages}{1120}
  (\bibinfo{year}{2015}).

\bibitem[{\citenamefont{Gao et~al.}(2016)\citenamefont{Gao, Zhang, Zhang, Wang,
  Huang, and Xia}}]{Gao_SciRep_2016}
\bibinfo{author}{\bibfnamefont{G.}~\bibnamefont{Gao}},
  \bibinfo{author}{\bibfnamefont{Y.}~\bibnamefont{Zhang}},
  \bibinfo{author}{\bibfnamefont{H.}~\bibnamefont{Zhang}},
  \bibinfo{author}{\bibfnamefont{Y.}~\bibnamefont{Wang}},
  \bibinfo{author}{\bibfnamefont{Q.}~\bibnamefont{Huang}}, \bibnamefont{and}
  \bibinfo{author}{\bibfnamefont{J.}~\bibnamefont{Xia}}, \bibinfo{journal}{Sci.
  Rep.} \textbf{\bibinfo{volume}{6}}, \bibinfo{pages}{19999}
  (\bibinfo{year}{2016}).

\bibitem[{\citenamefont{McGarvey-Lechable
  et~al.}(2017)\citenamefont{McGarvey-Lechable, Hamidfar, Patel, Xu, Plant, and
  Bianucci}}]{KML_OE_2017}
\bibinfo{author}{\bibfnamefont{K.}~\bibnamefont{McGarvey-Lechable}},
  \bibinfo{author}{\bibfnamefont{T.}~\bibnamefont{Hamidfar}},
  \bibinfo{author}{\bibfnamefont{D.}~\bibnamefont{Patel}},
  \bibinfo{author}{\bibfnamefont{L.}~\bibnamefont{Xu}},
  \bibinfo{author}{\bibfnamefont{D.~V.} \bibnamefont{Plant}}, \bibnamefont{and}
  \bibinfo{author}{\bibfnamefont{P.}~\bibnamefont{Bianucci}},
  \bibinfo{journal}{Opt. Express} \textbf{\bibinfo{volume}{25}},
  \bibinfo{pages}{3916} (\bibinfo{year}{2017}).

\bibitem[{\citenamefont{Lo et~al.}(2018)\citenamefont{Lo, Lee, Weiss, and
  Fauchet}}]{Lo_OL_2018}
\bibinfo{author}{\bibfnamefont{S.~M.} \bibnamefont{Lo}},
  \bibinfo{author}{\bibfnamefont{J.~Y.} \bibnamefont{Lee}},
  \bibinfo{author}{\bibfnamefont{S.~M.} \bibnamefont{Weiss}}, \bibnamefont{and}
  \bibinfo{author}{\bibfnamefont{P.~M.} \bibnamefont{Fauchet}},
  \bibinfo{journal}{Opt. Lett.} \textbf{\bibinfo{volume}{43}},
  \bibinfo{pages}{2957} (\bibinfo{year}{2018}).

\bibitem[{\citenamefont{Johnson and Joannopoulos}(2001)}]{Johnson_OE_2001}
\bibinfo{author}{\bibfnamefont{S.~G.} \bibnamefont{Johnson}} \bibnamefont{and}
  \bibinfo{author}{\bibfnamefont{J.~D.} \bibnamefont{Joannopoulos}},
  \bibinfo{journal}{Opt. Express} \textbf{\bibinfo{volume}{8}},
  \bibinfo{pages}{173} (\bibinfo{year}{2001}).

\bibitem[{\citenamefont{Morichetti et~al.}(2010)\citenamefont{Morichetti,
  Canciamilla, Ferrari, Torregiani, Melloni, and
  Martinelli}}]{morichetti2010roughness}
\bibinfo{author}{\bibfnamefont{F.}~\bibnamefont{Morichetti}},
  \bibinfo{author}{\bibfnamefont{A.}~\bibnamefont{Canciamilla}},
  \bibinfo{author}{\bibfnamefont{C.}~\bibnamefont{Ferrari}},
  \bibinfo{author}{\bibfnamefont{M.}~\bibnamefont{Torregiani}},
  \bibinfo{author}{\bibfnamefont{A.}~\bibnamefont{Melloni}}, \bibnamefont{and}
  \bibinfo{author}{\bibfnamefont{M.}~\bibnamefont{Martinelli}},
  \bibinfo{journal}{Physical Review Letters} \textbf{\bibinfo{volume}{104}},
  \bibinfo{pages}{033902} (\bibinfo{year}{2010}).

\bibitem[{\citenamefont{Moss et~al.}(2013)\citenamefont{Moss, Morandotti,
  Gaeta, and Lipson}}]{Moss_NatPhoton_2013}
\bibinfo{author}{\bibfnamefont{D.~J.} \bibnamefont{Moss}},
  \bibinfo{author}{\bibfnamefont{R.}~\bibnamefont{Morandotti}},
  \bibinfo{author}{\bibfnamefont{A.~L.} \bibnamefont{Gaeta}}, \bibnamefont{and}
  \bibinfo{author}{\bibfnamefont{M.}~\bibnamefont{Lipson}},
  \bibinfo{journal}{Nat. Photon.} \textbf{\bibinfo{volume}{7}},
  \bibinfo{pages}{597} (\bibinfo{year}{2013}).

\bibitem[{\citenamefont{O'Brien et~al.}(2009)\citenamefont{O'Brien, Furusawa,
  and Vučković}}]{Obrien_NatPhoton_2009}
\bibinfo{author}{\bibfnamefont{J.~L.} \bibnamefont{O'Brien}},
  \bibinfo{author}{\bibfnamefont{A.}~\bibnamefont{Furusawa}}, \bibnamefont{and}
  \bibinfo{author}{\bibfnamefont{J.}~\bibnamefont{Vučković}},
  \bibinfo{journal}{Nat. Photon.} \textbf{\bibinfo{volume}{3}},
  \bibinfo{pages}{687} (\bibinfo{year}{2009}).

\bibitem[{\citenamefont{Wilczek}(1982{\natexlab{a}})}]{Wilczek_PRL_1982}
\bibinfo{author}{\bibfnamefont{F.}~\bibnamefont{Wilczek}},
  \bibinfo{journal}{Phys. Rev. Lett.} \textbf{\bibinfo{volume}{49}},
  \bibinfo{pages}{957} (\bibinfo{year}{1982}{\natexlab{a}}).

\bibitem[{\citenamefont{Wilczek}(1982{\natexlab{b}})}]{wilczek1982magnetic}
\bibinfo{author}{\bibfnamefont{F.}~\bibnamefont{Wilczek}},
  \bibinfo{journal}{Phys. Rev. Lett.} \textbf{\bibinfo{volume}{48}},
  \bibinfo{pages}{1144} (\bibinfo{year}{1982}{\natexlab{b}}).

\bibitem[{\citenamefont{Molchan et~al.}(2008)\citenamefont{Molchan, Doktorov,
  and Vlasov}}]{Molchan_JOA_2008}
\bibinfo{author}{\bibfnamefont{M.~A.} \bibnamefont{Molchan}},
  \bibinfo{author}{\bibfnamefont{E.~V.} \bibnamefont{Doktorov}},
  \bibnamefont{and} \bibinfo{author}{\bibfnamefont{R.~A.}
  \bibnamefont{Vlasov}}, \bibinfo{journal}{J. Opt. A: Pure Appl. Opt.}
  \textbf{\bibinfo{volume}{11}}, \bibinfo{pages}{015706}
  (\bibinfo{year}{2008}).

\bibitem[{\citenamefont{Jesus-Silva et~al.}(2012)\citenamefont{Jesus-Silva,
  Fonseca, and Hickmann}}]{Jesus-Silva_OL_2012}
\bibinfo{author}{\bibfnamefont{A.~J.} \bibnamefont{Jesus-Silva}},
  \bibinfo{author}{\bibfnamefont{E.~J.~S.} \bibnamefont{Fonseca}},
  \bibnamefont{and} \bibinfo{author}{\bibfnamefont{J.~M.}
  \bibnamefont{Hickmann}}, \bibinfo{journal}{Opt. Lett.}
  \textbf{\bibinfo{volume}{37}}, \bibinfo{pages}{4552} (\bibinfo{year}{2012}).

\bibitem[{\citenamefont{Martinez-Castellanos and
  Guti{\'{e}}rrez-Vega}(2013)}]{Martinez-Castellanos_JOSAA_2013}
\bibinfo{author}{\bibfnamefont{I.}~\bibnamefont{Martinez-Castellanos}}
  \bibnamefont{and} \bibinfo{author}{\bibfnamefont{J.~C.}
  \bibnamefont{Guti{\'{e}}rrez-Vega}}, \bibinfo{journal}{J. Opt. Soc. Am. A}
  \textbf{\bibinfo{volume}{30}}, \bibinfo{pages}{2395} (\bibinfo{year}{2013}).

\bibitem[{\citenamefont{Martinez-Castellanos and
  Guti{\'{e}}rrez-Vega}(2015)}]{Martinez-Castellanos_OL_2015}
\bibinfo{author}{\bibfnamefont{I.}~\bibnamefont{Martinez-Castellanos}}
  \bibnamefont{and} \bibinfo{author}{\bibfnamefont{J.~C.}
  \bibnamefont{Guti{\'{e}}rrez-Vega}}, \bibinfo{journal}{Opt. Lett.}
  \textbf{\bibinfo{volume}{40}}, \bibinfo{pages}{1764} (\bibinfo{year}{2015}).

\bibitem[{\citenamefont{Wang et~al.}(2016)\citenamefont{Wang, Zhao, Feng, Xu,
  Liu, Cui, Zhang, and Huang}}]{Wang_SR_2016}
\bibinfo{author}{\bibfnamefont{Y.}~\bibnamefont{Wang}},
  \bibinfo{author}{\bibfnamefont{P.}~\bibnamefont{Zhao}},
  \bibinfo{author}{\bibfnamefont{X.}~\bibnamefont{Feng}},
  \bibinfo{author}{\bibfnamefont{Y.}~\bibnamefont{Xu}},
  \bibinfo{author}{\bibfnamefont{F.}~\bibnamefont{Liu}},
  \bibinfo{author}{\bibfnamefont{K.}~\bibnamefont{Cui}},
  \bibinfo{author}{\bibfnamefont{W.}~\bibnamefont{Zhang}}, \bibnamefont{and}
  \bibinfo{author}{\bibfnamefont{Y.}~\bibnamefont{Huang}},
  \bibinfo{journal}{Sci. Rep.} \textbf{\bibinfo{volume}{6}},
  \bibinfo{pages}{36269} (\bibinfo{year}{2016}).

\bibitem[{\citenamefont{Pisanty et~al.}(2019)\citenamefont{Pisanty, Machado,
  Vicu{\~{n}}a-Hern{\'{a}}ndez, Pic{\'{o}}n, Celi, Torres, and
  Lewenstein}}]{Pisanty_NatPhoton_2019}
\bibinfo{author}{\bibfnamefont{E.}~\bibnamefont{Pisanty}},
  \bibinfo{author}{\bibfnamefont{G.~J.} \bibnamefont{Machado}},
  \bibinfo{author}{\bibfnamefont{V.}~\bibnamefont{Vicu{\~{n}}a-Hern{\'{a}}ndez}},
  \bibinfo{author}{\bibfnamefont{A.}~\bibnamefont{Pic{\'{o}}n}},
  \bibinfo{author}{\bibfnamefont{A.}~\bibnamefont{Celi}},
  \bibinfo{author}{\bibfnamefont{J.~P.} \bibnamefont{Torres}},
  \bibnamefont{and}
  \bibinfo{author}{\bibfnamefont{M.}~\bibnamefont{Lewenstein}},
  \bibinfo{journal}{Nat. Photon.} \textbf{\bibinfo{volume}{13}},
  \bibinfo{pages}{569} (\bibinfo{year}{2019}).

\end{thebibliography}

\end{document}